\documentclass[fleqn,usenatbib, twocolumn]{mnras}
\usepackage[T1]{fontenc}
\usepackage{ae,aecompl}
\usepackage{rotating}
\usepackage{caption}
\usepackage{graphicx}
\usepackage{longtable}
\usepackage{xcolor}
  \usepackage{ulem}
\usepackage{graphicx}	% Including figure files
\usepackage{amsmath}	% Advanced maths commands
\usepackage{amssymb}	% Extra maths symbols
\usepackage{color}
\usepackage{url}
\usepackage{comment}
\title[]{Long-term evolutions of the cyclotron line energies in Her X-1, Vela X-1 and Cen X-3 as observed with {\it Swift}/BAT}

\author[L., Ji et al.]{
	L. Ji$^{1}$\thanks{E-mail: ji.long@astro.uni-tuebingen.de},
	R. Staubert$^{1}$\thanks{E-mail: staubert@astro.uni-tuebingen.de},
	L. Ducci$^{1}$\thanks{E-mail: ducci@astro.uni-tuebingen.de},
	A. Santangelo$^{1}$\thanks{E-mail: andrea.santangelo@uni-tuebingen.de} 
	S. Zhang$^{2}$\thanks{E-mail: szhang@ihep.ac.cn}
	Z. Chang$^{2}$ \\
	$^{1}$ Institut f\"ur Astronomie und Astrophysik, Kepler Center for Astro and Particle Physics, Sand 1, D-72076 T\"ubingen, Germany\\
	$^{2}$Key Laboratory for Particle Astrophysics, Institute of High Energy Physics, Beijing 100049, China\\
	}

\date{Accepted XXX. Received YYY; in original form ZZZ}

\pubyear{2019}

\begin{document}
\label{firstpage}
\pagerange{\pageref{firstpage}--\pageref{lastpage}}
\maketitle

\begin{abstract}
We study the long-term evolution of the centroid energy of cyclotron lines - often referered to as Cyclotron Resonance Scattering Features (CRSF) - in Her X-1, Vela X-1 and Cen X-3, using survey observations of the Burst Alert Telescope (BAT) onboard {\it Swift}. 
We find a significant decrease of the fundamental CRSF energy in Her X-1 and the first harmonic line energy in Vela X-1, since the launch of {\it Swift} in 2004 and until 2010 and 2012, respectively.
In both sources the decreases stopped at some time, with a quite stable centroid energy thereafter.
Unlike in Her X-1 and Vela X-1, the CRSF energy in Cen X-3 does not show a long-term decrease. 
It is observed not to change for at least the past 14 years. 
The long-term variation of the line energy is a direct way to investigate the magnetic field structure in the polar regions of pulsars. 
Our results may stimulate the development of theoretical models, especially regarding to how the accreted mass accumulates in the accretion  mound or how the magnetic field distorts around the polar cap.

\end{abstract}

%%%%%%%%%%%%%%%%%%%%%%%%%%%%%%%%%%%%%%%%%%%%%%%%%%

%%%%%%%%%%%%%%%%% BODY OF PAPER %%%%%%%%%%%%%%%%%%
\section{Introduction}
Accretion powered  X-ray pulsars are some of the brightest sources in our Galaxy. Their magnetic fields are believed to be of the order of $\sim$ $10^{11}$-$10^{13}$\,G. 
The accreting matter is funnelled from the magnetospheric radius to a small region on the surface of the neutron star \citep[the polar cap, see, e.g.,][]{Basko1976M}. 
In the presence of a strong magnetic field, the energies of the electrons with respect to their movement perpendicular to the magnetic field are quantized into discrete Landau levels. 
Resonant scattering of photons on such electrons in the line-forming region,  results in cyclotron resonance scattering features (CRSFs) simply referred to as cyclotron lines. 
The centroid cyclotron line energy is $E_{\rm cyc}$ = $11.6nB_{12}(1+z)^{-1}$, where the $B_{12}$ is the magnetic field strength in units of $10^{12}$ Gauss and $z$ is the gravitational redshift in the line-forming region. n=1 and n=2,3,4... correspond to the fundamental and harmonic cyclotron lines, respectively \footnote{In this paper, we quote the fundamental and the first harmonic as $E_{\rm cyc}$ and $E_{\rm cyc\_H}$, respectively.}. 
Cyclotron lines provide a direct measure of the magnetic field strength in the line-forming region, and its variability reflects the changes of the accretion geometry and/or the re-arrangement of the magnetic field configuration \citep[see, e.g.,][]{Becker2012, Mushtukov2015}.

It has been found that the CRSF energy generally depends on pulse phase, often on luminosity \citep[see, e.g., ][]{Staubert2007, Klochkov2011,Vasco2013, Furst2014, Vybornov2017}.
In addition, the long-term time dependence was discovered by \citet{Staubert2014, Staubert2016} in Her X-1, in which the cyclotron line energy decreases by $\sim$ 5\,keV over 20 years.
They used the data obtained with several X-ray observatories ({\it RXTE}, {\it Beppo-SAX}, {\it INTEGRAL}, {\it Suzaku} and {\it NuSTAR}), in the time period from 1996 to 2015. 
Subsequently, \citet{Klochkov2015} independently confirmed this result by using monitoring {\it Swift}/BAT observations. 
What is more interesting is that recently \citet{Staubert2017} proposed that the 20-year $E_{\rm cyc}$ decrease has ended and an inverse trend could start soon. 
In this paper, {we have started} a detailed re-analysis of {\it Swift}/BAT data extending to the most recent observations, in order to follow the evolution of the cyclotron  line energies over time. 
In addition to Her X-1, a second source was found - Vela X-1 - showing a similar long-term decrease of its first harmonic cyclotron line energy \citep{LaParola2016}. {\it Swift}/BAT data of Vela X-1 were analysed with the software {\sc bat\_imager} \citep{Segreto2010}, and it was found that the first harmonic cyclotron line decreased by $\sim$ 0.72 ${\rm keV\,yr^{-1}}$ between December 2004 and June 2010, and then remained constant.
Additionally, there are two other candidates showing CRSF variations with time: V0332+53 and 4U 1538-22. 
The former shows, in its 2015 outburst, a systematically lower cyclotron line energy in the declining phase of the outburst compared to the rising phase for equal levels of luminosity \citep{Cusumano2016, Doroshenko2017, Vybornov2017}.
For 4U 1538-22 a possible increase by $\sim$1.5\,keV may have happened between the {\it RXTE} and {\it Suzaku} observations, which are about $\sim$ 8.5 years apart \citep{Hemphill2016}.

In this work, we re-analyze {\it Swift}/BAT monitoring observations of Her X-1 and Vela X-1 by using a procedure and software developed by \citet{Klochkov2015}, which had led to the confirmation of the long-term decrease of the cyclotron line energy in Her X-1. We improve previous studies in four aspects: \\
1) Data after 2015, especially in Her X-1 when the decrease trend ended, are included. \\
2) The flux-correction (see below) is taken into account in Her X-1. \\
3) Updated calibration files are employed, which improves the results for all sources. \\
4) In addition to Her X-1 and Vela X-1, we searched for a long-term variation of cyclotron line energies in
other sources by using archived {\it Swift}/BAT data. We considered all the source in Table~1 published by \citet{Maitra2017},
however, only in Cen X-3 {was it possible to detect the cyclotron line \citep[see, e.g., ][]{Santangelo1998} with sufficient significance.}

Therefore, in this paper we present the evolution of the cyclotron line energies in Her X-1, Vela X-1 and Cen X-3. 
The paper is organized as follows: the detailed data reduction and the corresponding results are shown in Section 2 
and 3, respectively. 
We discuss the implication of our results in Section 4.

% Fig. 1
\begin{figure}
	\centering
	\includegraphics[width=3.2in]{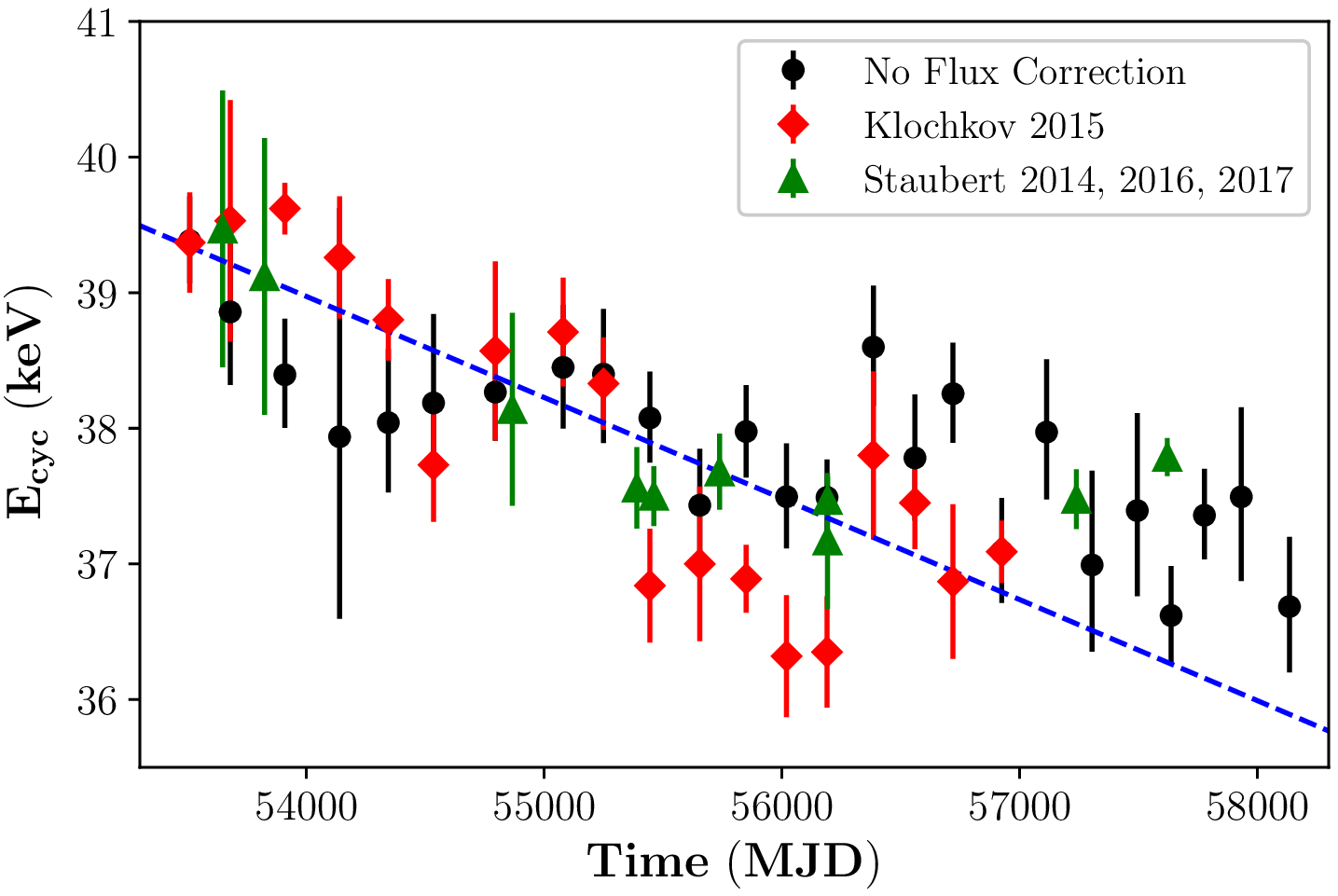}
	\caption{The long-term evolution of the CRSF energy in Her X-1 observed by	{\it Swift}/BAT. The black and red points are the results  in this paper (without the flux correction, see text) and previous results reported by \citet{Klochkov2015}, respectively.	
	 The green points are taken from \citet{Staubert2014, Staubert2016, Staubert2017}, which are observed with {\it RXTE}, {\it INTEGRAL}, {\it Suzaku}
	and {\it NuSTAR}. The blue dashed line is taken from \citet{Staubert2014} as well, which represents the long-term decrease of the centroid energy of the CRSF. 
}
	\label{fig_her}
\end{figure}

\section{OBSERVATIONS AND DATA ANALYSIS}
{\it Swift}/BAT is a coded aperture telescope operating in the 15--150 keV range \citep{Barthelmy2005}. 
The data we selected have been taken in the {\it survey} mode, for which events were collected in the detector plane histograms (DPHs), typically with a five-minute exposure time. 
All data available since the launch of the mission in 2004 have been used.
In this paper, we generally followed the data reduction of \citet{Klochkov2015}. Here we briefly summarize the procedures. 
We reconstructed the sky map in each observation with the tools {\it "batbinevt"} and {\it "batfftimage"} from the {\sc heasoft ver. 6.21}  package \footnote{\url{https://heasarc.gsfc.nasa.gov/docs/software/lheasoft/}}.
We extracted the spectra only if the source could be identified in the sky map. 
We used the "{\it beterebin}" tool to correct the gain/offset of detectors with the latest CALDB that was released in October 2017. 
As suggested by the BAT team, we added the energy-dependent systematic errors by using the {\it "batphasyserr"} tool. 
In order to model the cyclotron line, we used a Gaussian absorption line ({\it "gabs"} in {\sc xspec}, i.e., 
$exp\left\{-\frac{\rm Depth}{\sqrt{2\pi} \sigma} e^{-\frac{1}{2} \frac{(E-E_{\rm cyc})^2}{{\sigma}^2}}\right\}$).
This model has been widely used to describe cyclotron lines \citep[e.g.,][]{Staubert2007, Furst2013}.
For the continuum, different functions were used for different sources (see below). 
The energy band used for the fits is 15-70\,keV. 
To enhance the statistics in the spectral analysis, we jointly fitted tens to hundreds of spectra in a given time interval, for which the spectral shapes were assumed to be the same while normalizations were variable. 
The validity of this method has been verified by \citet{Klochkov2015}.
In this paper, all uncertainties quoted correspond to a 68\% confidence level.

\begin{figure}
    \centering
    \includegraphics[width=3.2in]{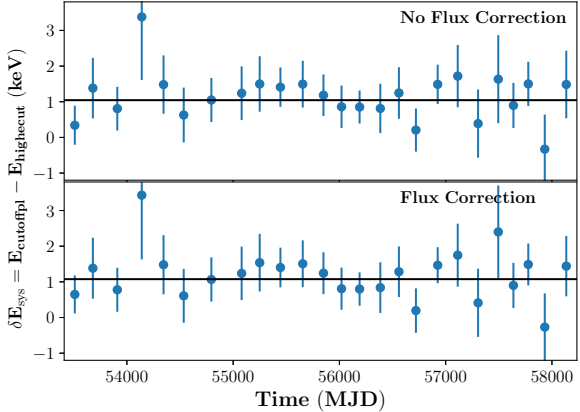}
    \caption{
        {
            The systematic difference ($\delta E_{\rm sys}$) of the detected CRSF centroid energy by using different continuum models, i.e., {\it highecut*powerlaw} and {\it cutoffpl}.
            On average the latter results in a higher $E_{\rm cyc}$ by 1.08$\pm$ 0.10 (1.04$\pm$ 0.10)\,keV if the flux correction is (not) considered.
        }
    }.
    \label{compare}
\end{figure}

% Fig. 2
\begin{figure}
	\centering
	\includegraphics[width=3.2in]{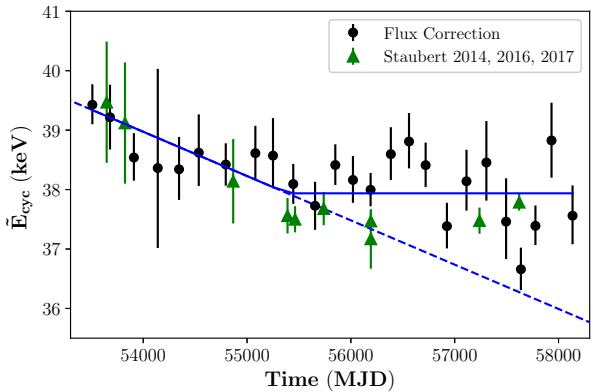}
	\caption{
        The long-term evolution of the CRSF energy in Her X-1 observed by	{\it Swift}/BAT. 
        The black points are the results of this analysis (including the flux correction, see text). The green points are taken from \citet{Staubert2014, Staubert2016, Staubert2017}, which are observed with {\it RXTE}, {\it INTEGRAL}, {\it Suzaku} and {\it NuSTAR}. 
        The blue dashed line is taken from \citet{Staubert2014}, which represents the long-term decrease of the centroid energy of the CRSF. 
    	After the end of the decrease, the mean cyclotron line energy seems to be constant at $E_{\rm cyc}$=37.94$\pm$0.12\,keV. 
}
	\label{fig_her_cor}
\end{figure}

\section{RESULTS}
\subsection{Her X-1}
We summed the BAT spectra into time intervals based on Table~1 in \citet{Klochkov2015}, and extended to  new  observations. 
We fitted the spectra with a {\it highecut} \footnote{\url{https://heasarc.gsfc.nasa.gov/xanadu/xspec/manual/node238.html}} model, i.e., a power-law  continuum with an exponential cut-off, modified by the CRSF component ({\it gabs}). 
This continuum model has been widely used for the long-term CRSF evolution \citep[see, e.g.,][]{Staubert2014}.
Thanks to the stable continuum shape of Her X-1, we froze the parameters at the e-folding energy $E_{\rm fold}$ =10\,keV, the cutoff energy $E_{\rm cut}$=21\,keV, and the powerlaw index $\Gamma$ = 0.9 \citep{Furst2013}.
We show the best-fitting CRSF energy values in Figure~\ref{fig_her}, together with previous results of  \citet{Klochkov2015} and \citet{Staubert2014, Staubert2016, Staubert2017}. 
We found that the line energy in this paper was  systematically different from that reported by \citet{Klochkov2015}, although the same data and fitting method were used. We note that there are two main reasons:
{
1) the continuum models have a slight influence on the line detection. 
We fitted the spectra with two continuum models, i.e., the {\it highecut} and the {\it cutoffpl} (a power law with high energy exponential rolloff) used by \citet{Klochkov2015}, and found that on average the latter resulted in a systematically higher $E_{\rm cyc}$ by $\delta E_{sys}$ = 1.04 $\pm$ 0.10\,keV. 
We show the comparison in Figure~\ref{compare}.
2) we used the updated detector gain calibration of {\it Swift}/BAT in this paper, which had $\sim$ 4\% gain shift  during 2004-2011. Therefore, the resulting shift of the cyclotron line energy is $\delta {\rm gain}\times E_{\rm cyc}$.
We compared the expected shift with observations by using a $\chi^{2}$-test, which leads to a reduced-$\chi^2$ of 1.2 (18 dof) with a p-value of 0.25.
This suggests that the CRSF energy reported in this paper is well in agreement with that detected by \citet{Klochkov2015} after considering the above discussed two effects. 
}
In addition, it is worth noting that the CRSF energy starts to significantly deviate from the downwards trend since $\sim$ MJD 56500, which is consistent with {\it NuSTAR} observations, i.e., the last two green points in Figure~\ref{fig_her}.

It is well-known that in Her X-1 the CRSF energy is related to the luminosity ($L_{\rm x}$) \citep{Staubert2007, Staubert2016}.
Therefore, in the following spectral analysis we took the $E_{\rm cyc}/L{\rm x}$ correlation into account, by using the maximum flux of the individual 35d Main-On cycle from which the data were taken as a reference.
Historically, the maximum Main-On flux, as measured by {\it RXTE}/ASM in the energy range of 2--10\,keV, was taken as a measure of $L_{\rm x}$ of this particular 35d cycle \citep[see, e.g.,][]{Staubert2007}.
In this paper we follow this approach, using flux measurements by {\it Swift}/BAT, which are converted into units of ASM-cts/s according to the following formula: (2-10\,keV ASM-cts/s) = 93 $\times$ (15-50\,keV BAT-cts/s\,$\rm {cm}^{-2}\,s^{-1}$) \citep[for details see Appendix A.2. in ][]{Staubert2016}. 
The $E_{\rm cyc}$/luminosity correlation is $\delta E_{\rm cyc}$ $\propto$ $f\times \delta$ Flux, where $f$ is a scaling factor allowing to normalize the measured cyclotron line energy to a reference flux. 
Following \citet{Staubert2014}, the reference flux is $\rm Flux_{ref}$ = 6.8 ASM-cts/s, where the CRSF energy is assumed to be $E_{\rm ref}$.
For data until 2012 {(i.e., MJD 55927)} we used the scaling factor 0.44\,keV/ASM-cts/s as stated by \citet{Staubert2016,Staubert2017}. 
For data after 2012 we used a scaling factor of 0.70\,keV/ASM-cts/s, which was found to describe the data from 2012 to February 2018 with high precision (private communication with Staubert).

Generally, one point shown in Figure \ref{fig_her} comprises tens to hundreds of spectra from several 35d cycles. 
Since different 35d cycles have different maximum Main-On fluxes, the combined fitting of those spectra was done by applying appropriate scaling factors to spectra from different 35d cycles (which can easily be done within the XSPEC fitting software).
So, for a spectrum ($S_{\rm i}$), its CRSF energy  could be written as   $E_{\rm cyc\_i}$ = $E_{\rm ref}$ + ($\rm Flux_{\rm i} - Flux_{ref} $)$\times f$, where the $\rm Flux_{\rm i}$ is the maximum flux of the corresponding 35d Main-On. 
The  $E_{\rm ref}$ can be regarded as a fitting parameter in the spectral analysis. 
In this way we have obtained the long-term evolution of the cyclotron line energy for the reference flux ($\rm Flux_{ref}$ = 6.8 ASM-cts/s), i.e., the flux-corrected $\tilde E_{\rm cyc}$. We show the result in Fig.~\ref{fig_her_cor}.
For comparison, we also include the line energies measured by other satellites as reported by \citet{Staubert2014, 
Staubert2016, Staubert2017}.
It is evident that the line energy decreases until a certain time $t_{\rm crit}$, which is in good agreement with previous reports (the blue dashed line). After that, the line energy significantly deviates from the linear decreasing trend, and in general remains unchanged. 
We note that an alternative continuum model {\it cutoffpl} only leads to a systematically higher line energy by 1.08 $\pm$ 0.10 (Figure~\ref{compare}), and does not have an influence on the trend.
We used a break line to fit the {flux-corrected $\tilde E_{\rm cyc}$} evolution of {\it Swift} observations as \begin{eqnarray}
{\tilde E_{\rm cyc}}(t) = \left\{ \begin{array}{l}
{E_{\rm 0}} + b \times (t - {t_{\rm 0}}),t \le {t_{\rm crit}}\\
const = {E_{\rm 0}} + a \times ({t_{\rm crit}} - {t_{\rm 0}}),t > {t_{\rm crit}}
\end{array} \right.
\end{eqnarray}
where  $t_{\rm 0}$=MJD~53500, $E_{\rm 0}$=39.88\,keV and b=$\rm -7.22\times {10}^{-4}\,keV/d$ by following \citet{Staubert2014}.
The resulting  $t_{\rm crit}$ is MJD 55400$\pm$ 200, and the corresponding $\tilde E_{\rm cyc}$ after $t_{\rm crit}$ is at  $37.94\pm0.12$\,keV, showing as the horizontal blue line in Figure~\ref{fig_her_cor}.
We note, that this value is systematically higher (by 0.32\,keV) than the mean of {\it NuSTAR} and {\it Suzaku} observations (i.e., the green points) after $t_{\rm crit}$ \footnote{The comparison of the blue and the green points in Fig.~\ref{fig_her_cor} later that MJD 55400 may point to a calibration issue between {\it NuSTAR}/{\it Suzaku} and {\it Swift}/BAT}.

\subsection{Vela X-1}
\begin{figure}
	\centering
	\includegraphics[width=3.2in]{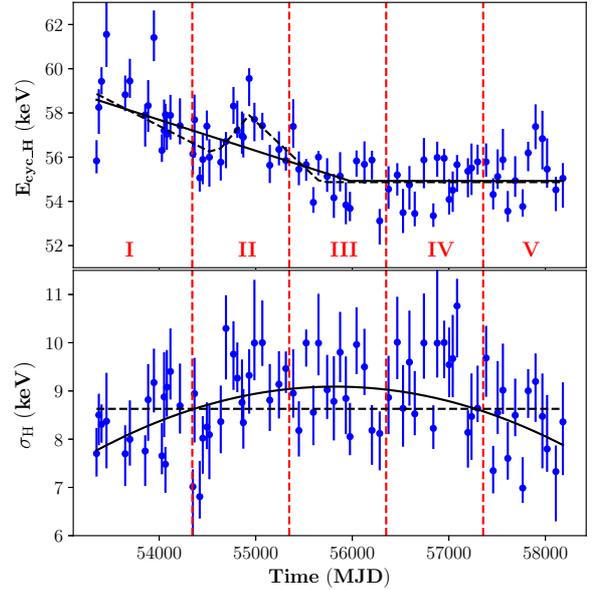}
	\caption{
Upper panel: the long-term evolution of the centroid energy of the harmonic CRSF in Vela X-1 observed with {\it Swift}/BAT. 
For the sake of comparison, we divided the observations into five epochs, the first four of which were defined in \citet{LaParola2016}. 
The line energy decreases significantly in the first two or three epochs with a rate of $-0.51\pm0.09$\,keV. 
After that, it remains almost unchanged. 
In addition, there seems to be a hump around the time MJD $\sim$ 55000, although the corresponding significance is only 1.9 $\sigma$ by comparing the solid and dashed black lines. 
Bottom panel: the long-term evolution of the CRSF width. 
We fitted the points with a constant line (dashed line) and a quadratic function (solid line), respectively. 
We found that the latter resulted in a better fitting at 3.5\,$\sigma$ confidence level, which hinted that the width might be variable as well besides the
$E_{\rm cyc\_H}$.}
\label{fig_vela}
\end{figure}

Following \citet{LaParola2016}, we employed a Comptonization model ({\it compTT} in {\sc xspec}) to describe the continuum of Vela X-1. 
We note that the fundamental line around 28\,keV cannot be detected by BAT in Vela X-1, and in this paper we only concentrate on the first harmonic $E_{\rm cyc\_H}$. 
Following the above procedures, we extracted the spectra of Vela X-1, and did the spectral analysis. 
Although observed by \citet{LaParola2016} and \citet{Furst2013} that the $E_{\rm cyc\_H}$ is related to the luminosity  in Vela X-1, their relation has not been well constrained yet (other than in Her X-1). 
Therefore, in the following analysis, we did not consider the flux-correction. 
Actually, as shown above, the flux-correction has little influence on the trend of the cyclotron line evolution detected by BAT because the stochastic variability of the luminosity is expected to be mitigated. 
During the spectral analysis, we fixed the temperature of seed photons at $kT_{\rm seed}$ = 1\,keV, because of no energy coverage below 15\,keV for BAT. 
We show the results in Table~\ref{tab_vela} and Figure~\ref{fig_vela}. 
For the sake of comparison, we divided the observations into five epochs, the first four of which were already defined in \citet{LaParola2016}. 
The line energy decreased significantly in the first two epochs, remaining almost unchanged thereafter.  
Therefore, we tried to fit the line evolution with a piecewise function as mentioned above. 
The critical point is around MJD~55980  (February 2012). 
Before MJD 55980, the decrease rate of $E_{\rm cyc\_H}$ is -0.51 $\pm$ 0.09\,keV per year. 
The $E_{\rm cyc\_H}$ after MJD 55980 is 54.96 $\pm$ 0.19\,keV. 
In addition, it seems that there is a hump around MJD  $\sim$ 55000. 
We fitted the $E_{\rm cyc\_H}$ variation with a multi-segment function (the black dashed line in Figure~\ref{fig_vela}), however, the confidence level of the presence of a hump is only at 1.9 $\sigma$, estimated by an F-test.
Apart from the $E_{\rm cyc\_H}$, we also found a hint that the width $\sigma_{\rm H}$ might be variable.
We fitted the $\sigma_{\rm H}$ evolution with a constant and a quadratic function, respectively.
An F-test shows that the latter is better at a 3.5$\sigma$ confidence level.
We confirmed that the source of the variability of $\sigma_{\rm H}$ is not instrumental because such a trend did not appear in other sources.

% Fig. 3

\subsection{Cen X-3 and other sources}
We tried to apply the above method in more sources. 
We considered the source list in Table~1 in \citet{Maitra2017} as a reference. 
We only considered persistent sources that could show the smooth long-term evolution of CRSFs. 
However, we found that most of sources (except for Cen X-3 and GX 301-2) could not be identified from the mosaic which was a prerequisite for extracting spectra, mainly because of the sensitivity of BAT. 
We found that the minimum flux to be detected in one DPH in the survey mode is approximately 0.02  $\rm cts\ s^{-1}\ {cm}^{-2} $, i.e., 90\,mCrab.
In GX 301-2, we could not constrain the cyclotron line, even if different continuum models were tested. 
This might be due to the dramatic changes of the CRSF in GX 301-2 with different orbital phases and pulse phases \citep[see, e.g., ][]{Kreykenbohm2004,LaBarbera2005, Suchy2012}, which might wipe out the absorption if stacking hundreds of spectra during fits.

\begin{figure}
	\centering
	\includegraphics[width=3.2in]{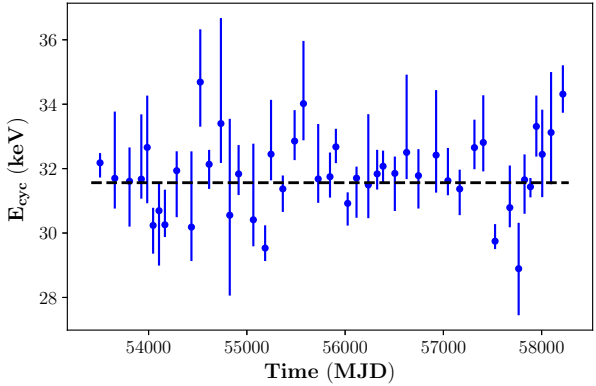}
	\caption{The long-term evolution of the CRSF in Cen X-3 observed with {\it Swift}/BAT.}
	\label{fig_cen}
\end{figure}

In Cen X-3, the cyclotron line can be constrained well. Here we used a Fermi-Dirac ({\it "fdcut"}) function to describe the continuum following \citet{Suchy2008}, where $fdcut(E)=AE^{-\Gamma} \frac{1}{1+e^{(E-E_{\rm cut})/E_{\rm fold}}}$ .
We show the best-fitting results in Table~\ref{tab_cen} and Figure~\ref{fig_cen}. Unlike the variability in Her X-1 and Vela X-1, we found that the centroid line energy in Cen X-3 was very stable over the past 14 years, at approximately 31.6$\pm$0.2\,keV.

% Fig. 4

\section{summary and discussion}
We searched for the long-term evolution of cyclotron line parameters in persistent sources by using archived {\it Swift}/BAT data. Because of the regular visiting and the large field-of-view, BAT provides nearly homogeneously spaced observations without long time gaps, which is a proper instrument for monitoring CRSFs in bright sources. In Her X-1, Vela X-1 and Cen X-3, we detected cyclotron lines well, and found significant line decreases in the first two sources.

For Her X-1, we confirmed the results of \citet{Klochkov2015}, while including two improvements: 1) We considered the flux-correction in the spectral analysis, which turned out to slightly influence the $E_{\rm cyc}$ values. 2) We analysed the data in the recent 4 years, which implied that the decrease of the CRSF centroid energy had ended around 2010 and remained {constant around} 37.94 $\pm$0.12\,keV afterwards. 
\citet{Staubert2016, Staubert2017} reported the continued decrease of the line energy until August 2015, when a particularly  low flux was measured. With the above mentioned change of the flux correction factor (to 0.7 instead of 0.44\,keV/ASM-cts/s) the flux corrected $\tilde E_{\rm cyc}$ for August 2015 is shifted upwards, leading to an estimate for the end of the decrease around  2012 to 2013 (Staubert et al., in preparation), which is consistent with the result derived here.
This, in turn, indicates that the long-term $\tilde E_{\rm cyc}$  decrease observed by BAT is robust because it is not so sensitive to the flux-correction  factor by averaging many spectra in different fluxes. In Vela X-1, we independently confirmed the decrease of the energy of the  first cyclotron harmonic reported by \citet{LaParola2016} by using different methods. The decrease rate is $0.51 \pm0.09$  $\rm keV\, yr^{-1}$, which is {similar to the decrease rate of $\sim$ 0.72\,$\rm keV\, yr^{-1}$  stated by \citet{LaParola2016}}. 
In addition to the centroid energy, we also found a long-term variation of the width of the cyclotron line, although the validity should be further confirmed by other observatories. 
We also searched for the cyclotron lines in other sources,
but only in Cen X-3 it is well detected. 
The line energy in Cen X-3 is very stable at least since 2004.

The variation of the cyclotron line energy (and width) observed in Her X-1 and Vela X-1 is believed to be caused by a local effect around the magnetic polar cap. 
As summarized by \citet{Staubert2014}, the variation might be due to a geometric displacement of the line-forming region, or a change of the local magnetic configuration. 
For example, the accreting matter accumulated in  the accretion mound, gradually resulting in a more extended line-forming region that corresponds to a lower  magnetic field. 
On the other hand, the magnetic field might be changed because of the drag by the accretion material \citep{Cheng1998, Zhang2006}, and the Hall drift and the Ohmic dissipation \citep{Goldreich1992}. 
However, as far as  we know, no model gives a conclusive explanation of the decrease of the cyclotron line energy. 
More observations and  theoretical works should be accumulated to investigate the complex magnetic field around the polar cap.

The time scale of the decrease of the line energy ($-E_{\rm cyc}$/$\dot{E_{\rm cyc}}$) in Her X-1/Vela X-1 is $\sim$ 100 years which is significantly shorter than the  characteristic time scale of the magnetic filed evolution in pulsars \citep{Bhattacharya1992}. 
Therefore, as suggested by \citet{Staubert2014}, the centroid energy of $E_{\rm cyc}$ may be cyclic on time scales of a few tens to hundreds of years, which may comprise of declining, stable and rising phases.
The results presented in this work show that BAT is a proper instrument to observe CRSFs in relatively bright sources such as Her X-1, Vela X-1, and Cen X-3, and supports the need for more sensitive and regular long-term monitoring of CRSFs in other accreting pulsars with observatories such as {\it NuSTAR}, {\it INTEGRAL}, {\it HXMT} and {\it Astrosat}.

%\clearpage

% former Fig. 2 after the end of the document

\section*{Acknowledgements}
JL thanks the support from the Chinese NSFC 11733009.
ZS thanks the support from XTP project XDA 04060604, the Strategic Priority Research Programme 'The Emergence of Cosmological Structures' of the Chinese Academy of Sciences, Grant No.XDB09000000,the National Key Research and Development Program of China (2016YFA0400800) and the Chinese NSFC 11473027 and 11733009.
We acknowledge discussions with Dr. Klochkov Dmitry, Dr. Markwardt Craig and Dr. Amy Lien, and help from the {\it Swift} Helpdesk.

\bibliographystyle{mnras}
\bibliography{mybibtex}

\appendix
\section{Best-fitting spectral parameters in Her X-1, Vela X-1 and Cen X-3}

\onecolumn
\begin{longtable}{ c|ccc|ccc }	
\caption{Best-fitting parameters of Her X-1 observed with {\it Swift}/BAT.  $E_{\rm cyc}$, $\sigma$ and Depth are the energy, width and depth of the CRSF.
The continuum is a powerlaw with a high energy cutoff ({\it highecut}), where the parameters are frozen at $\Gamma$=0.9, $E_{\rm cut}$=20.8 and
$E_{\rm fold}$=10.2 \citep{Furst2013}. During the spectral analysis, we considered the influence of the variable flux on the CRSF (the right three columns,
see text).}
\label{tab_her}\\
	\hline
	Time (MJD)  &  $E_{\rm cyc}$\,(keV) & $\sigma$\,(keV) & Depth & $E_{\rm cyc}$\,(keV) & $\sigma$\,(keV) & Depth \\
	\hline
	& \multicolumn{3}{|c|}{No Flux Correction} & \multicolumn{3}{|c}{Flux Correction}\\
	53440-53580 & $39.38_{-0.31}^{+0.33}$ & $4.04_{-0.24}^{+0.25}$ & $6.44_{-0.42}^{+0.44}$ & $39.43_{-0.33}^{+0.34}$ & $4.08_{-0.24}^{+0.26}$ & $6.15_{-0.42}^{+0.44}$ \\
	53610-53750 & $38.86_{-0.54}^{+0.55}$ & $4.11_{-0.38}^{+0.37}$ & $4.74_{-0.51}^{+0.52}$ & $39.22_{-0.54}^{+0.55}$ & $4.11_{-0.38}^{+0.37}$ & $4.74_{-0.51}^{+0.52}$ \\
	53790-54030 & $38.39_{-0.39}^{+0.41}$ & $4.06_{-0.25}^{+0.29}$ & $5.71_{-0.43}^{+0.47}$ & $38.54_{-0.39}^{+0.41}$ & $4.08_{-0.25}^{+0.29}$ & $5.72_{-0.43}^{+0.47}$ \\
	54070-54210 & $37.94_{-1.34}^{+1.69}$ & $4.30_{-0.97}^{+1.24}$ & $3.57_{-0.89}^{+1.14}$ & $38.36_{-1.35}^{+1.67}$ & $4.29_{-0.92}^{+1.21}$ & $3.56_{-0.87}^{+1.13}$ \\
	54240-54450 & $38.04_{-0.51}^{+0.54}$ & $3.45_{-0.38}^{+0.38}$ & $4.34_{-0.49}^{+0.52}$ & $38.34_{-0.51}^{+0.54}$ & $3.42_{-0.38}^{+0.39}$ & $4.33_{-0.49}^{+0.52}$ \\
	54480-54590 & $38.19_{-0.55}^{+0.66}$ & $4.26_{-0.30}^{+0.49}$ & $5.05_{-0.50}^{+0.65}$ & $38.62_{-0.56}^{+0.64}$ & $4.30_{-0.31}^{+0.47}$ & $5.08_{-0.51}^{+0.64}$ \\
	54620-54970 & $38.27_{-0.36}^{+0.36}$ & $3.77_{-0.25}^{+0.24}$ & $5.03_{-0.37}^{+0.38}$ & $38.42_{-0.36}^{+0.36}$ & $3.76_{-0.25}^{+0.24}$ & $5.02_{-0.37}^{+0.38}$ \\
	55010-55150 & $38.45_{-0.45}^{+0.46}$ & $4.12_{-0.32}^{+0.31}$ & $4.95_{-0.43}^{+0.44}$ & $38.61_{-0.45}^{+0.46}$ & $4.12_{-0.32}^{+0.31}$ & $4.94_{-0.43}^{+0.45}$ \\
	55180-55320 & $38.40_{-0.51}^{+0.48}$ & $4.17_{-0.37}^{+0.28}$ & $4.72_{-0.47}^{+0.44}$ & $38.57_{-0.55}^{+0.63}$ & $4.07_{-0.26}^{+0.41}$ & $4.66_{-0.40}^{+0.51}$ \\
	55360-55530 & $38.08_{-0.33}^{+0.34}$ & $3.97_{-0.23}^{+0.25}$ & $5.44_{-0.36}^{+0.38}$ & $38.09_{-0.33}^{+0.34}$ & $3.99_{-0.24}^{+0.23}$ & $5.42_{-0.36}^{+0.37}$ \\
	55570-55740 & $37.43_{-0.39}^{+0.42}$ & $4.00_{-0.25}^{+0.30}$ & $4.49_{-0.35}^{+0.38}$ & $37.73_{-0.40}^{+0.40}$ & $4.02_{-0.29}^{+0.27}$ & $4.49_{-0.36}^{+0.37}$ \\
	55780-55920 & $37.98_{-0.34}^{+0.34}$ & $3.98_{-0.24}^{+0.23}$ & $5.79_{-0.39}^{+0.40}$ & $38.41_{-0.34}^{+0.35}$ & $3.98_{-0.22}^{+0.25}$ & $5.80_{-0.38}^{+0.41}$ \\
	55950-56090 & $37.50_{-0.38}^{+0.39}$ & $3.90_{-0.28}^{+0.27}$ & $5.15_{-0.39}^{+0.41}$ & $38.16_{-0.38}^{+0.40}$ & $3.92_{-0.27}^{+0.29}$ & $5.13_{-0.39}^{+0.41}$ \\
	56120-56260 & $37.49_{-0.28}^{+0.28}$ & $4.04_{-0.20}^{+0.19}$ & $5.49_{-0.30}^{+0.30}$ & $38.00_{-0.28}^{+0.28}$ & $4.03_{-0.19}^{+0.21}$ & $5.47_{-0.30}^{+0.31}$ \\
	56300-56470 & $38.60_{-0.43}^{+0.45}$ & $4.57_{-0.28}^{+0.31}$ & $6.48_{-0.49}^{+0.53}$ & $38.60_{-0.43}^{+0.45}$ & $4.58_{-0.28}^{+0.31}$ & $6.48_{-0.49}^{+0.53}$ \\
	56510-56610 & $37.78_{-0.46}^{+0.47}$ & $3.80_{-0.34}^{+0.32}$ & $4.93_{-0.48}^{+0.49}$ & $38.81_{-0.46}^{+0.47}$ & $3.79_{-0.33}^{+0.34}$ & $4.90_{-0.47}^{+0.50}$ \\
	56650-56790 & $38.26_{-0.36}^{+0.38}$ & $4.42_{-0.23}^{+0.26}$ & $6.02_{-0.39}^{+0.42}$ & $38.41_{-0.37}^{+0.38}$ & $4.45_{-0.23}^{+0.26}$ & $6.02_{-0.39}^{+0.42}$ \\
	56820-57030 & $37.09_{-0.38}^{+0.40}$ & $4.04_{-0.28}^{+0.31}$ & $4.19_{-0.32}^{+0.34}$ & $37.38_{-0.37}^{+0.39}$ & $4.04_{-0.28}^{+0.30}$ & $4.20_{-0.32}^{+0.34}$ \\
	57060-57167 & $37.97_{-0.50}^{+0.54}$ & $4.53_{-0.32}^{+0.38}$ & $6.38_{-0.57}^{+0.65}$ & $38.14_{-0.51}^{+0.52}$ & $4.55_{-0.35}^{+0.35}$ & $6.39_{-0.60}^{+0.62}$ \\
	57200-57410 & $36.99_{-0.64}^{+0.70}$ & $3.20_{-0.49}^{+0.53}$ & $6.42_{-1.01}^{+1.13}$ & $38.46_{-0.64}^{+0.70}$ & $3.19_{-0.49}^{+0.53}$ & $6.39_{-1.00}^{+1.13}$ \\
	57440-57550 & $37.39_{-0.63}^{+0.72}$ & $4.10_{-0.49}^{+0.54}$ & $5.05_{-0.61}^{+0.68}$ & $37.46_{-0.63}^{+0.72}$ & $4.10_{-0.49}^{+0.55}$ & $4.95_{-0.59}^{+0.68}$ \\
	57585-57690 & $36.62_{-0.35}^{+0.36}$ & $3.52_{-0.28}^{+0.29}$ & $4.66_{-0.37}^{+0.39}$ & $36.66_{-0.35}^{+0.36}$ & $3.50_{-0.28}^{+0.29}$ & $4.66_{-0.37}^{+0.39}$ \\
	57725-57830 & $37.36_{-0.32}^{+0.34}$ & $3.96_{-0.24}^{+0.26}$ & $5.57_{-0.38}^{+0.40}$ & $37.39_{-0.32}^{+0.34}$ & $3.96_{-0.24}^{+0.26}$ & $5.56_{-0.38}^{+0.40}$ \\
	57865-58000 & $37.49_{-0.62}^{+0.66}$ & $4.24_{-0.39}^{+0.46}$ & $5.45_{-0.61}^{+0.69}$ & $38.83_{-0.63}^{+0.64}$ & $4.40_{-0.43}^{+0.41}$ & $5.71_{-0.65}^{+0.67}$ \\
	58030-58240 & $36.69_{-0.48}^{+0.51}$ & $3.81_{-0.33}^{+0.39}$ & $4.84_{-0.48}^{+0.52}$ & $37.56_{-0.48}^{+0.51}$ & $3.82_{-0.33}^{+0.39}$ & $4.82_{-0.47}^{+0.52}$ \\
	\hline
\end{longtable}

\clearpage
\begin{longtable}{ c|ccccc }
\caption{Best-fitting parameters of Vela X-1 observed with {\it Swift}/BAT. $E_{\rm cyc\_H}$\,(keV), $\sigma_{\rm H}\,(keV)$ and Depth$_{\rm H}$ are the energy, width and depth of the harmonic CRSF, respectively.  $kT_{\rm Comp}$ and $\tau_{\rm Comp}$ are the temperature and the optical depth of the Comptonized continuum spectra, respectively.}
\label{tab_vela}\\
		\hline
		Time (MJD) & $E_{\rm cyc\_H}$\,(keV) & $\sigma_{\rm H}\,(keV)$ & Depth$_{\rm H}$ & $kT_{\rm Comp}$\,(keV) &  $\tau_{\rm Comp}$\\
		\hline
		53351-53375 & $55.83_{-0.56}^{+0.93}$ & $7.70_{-0.48}^{+0.53}$ & $17.61_{-1.75}^{+2.86}$ & $6.41_{-0.10}^{+0.10}$ & $16.78_{-1.04}^{+1.14}$  \\
		53375-53403 & $58.25_{-1.20}^{+0.82}$ & $8.50_{-0.63}^{+0.44}$ & $22.05_{-3.01}^{+1.94}$ & $7.68_{-0.19}^{+0.18}$ & $10.64_{-0.53}^{+0.56}$  \\
		53403-53454 & $59.43_{-0.59}^{+0.64}$ & $8.31_{-0.38}^{+0.41}$ & $18.67_{-1.42}^{+1.60}$ & $6.78_{-0.04}^{+0.04}$ & $15.57_{-0.17}^{+0.16}$  \\
		53454-53588 & $61.55_{-1.47}^{+1.46}$ & $8.37_{-0.98}^{+1.00}$ & $20.43_{-4.42}^{+2.64}$ & $7.53_{-0.15}^{+0.16}$ & $10.94_{-0.61}^{+0.48}$  \\
		53648-53696 & $58.83_{-0.74}^{+0.87}$ & $7.70_{-0.67}^{+0.59}$ & $16.27_{-1.97}^{+2.08}$ & $7.26_{-0.12}^{+0.15}$ & $11.65_{-0.61}^{+0.55}$  \\
		53696-53855 & $59.44_{-0.76}^{+1.01}$ & $8.00_{-0.54}^{+0.81}$ & $18.08_{-0.75}^{+2.34}$ & $7.17_{-0.06}^{+0.15}$ & $12.43_{-0.75}^{+0.29}$  \\
		53855-53886 & $57.91_{-1.43}^{+0.49}$ & $7.75_{-0.72}^{+0.33}$ & $15.98_{-2.87}^{+1.63}$ & $6.85_{-0.09}^{+0.06}$ & $14.52_{-0.45}^{+0.66}$  \\
		53886-53945 & $58.33_{-1.19}^{+1.15}$ & $8.82_{-0.85}^{+0.74}$ & $22.68_{-4.51}^{+4.37}$ & $7.80_{-0.28}^{+0.20}$ & $10.17_{-0.45}^{+0.62}$  \\
		53945-53990 & $61.41_{-1.07}^{+1.22}$ & $9.17_{-0.62}^{+0.70}$ & $24.97_{-3.00}^{+3.70}$ & $7.09_{-0.06}^{+0.07}$ & $12.83_{-0.20}^{+0.18}$  \\
		54028-54051 & $56.30_{-0.52}^{+0.73}$ & $7.66_{-0.38}^{+0.68}$ & $16.09_{-1.81}^{+2.49}$ & $6.94_{-0.11}^{+0.14}$ & $13.70_{-0.82}^{+0.63}$  \\
		54051-54064 & $57.20_{-0.84}^{+1.41}$ & $8.88_{-0.83}^{+0.93}$ & $19.99_{-3.94}^{+3.82}$ & $7.33_{-0.21}^{+0.17}$ & $12.51_{-0.60}^{+0.88}$  \\
		54064-54079 & $57.93_{-0.90}^{+0.49}$ & $7.48_{-0.59}^{+0.36}$ & $19.10_{-2.51}^{+1.93}$ & $6.93_{-0.11}^{+0.13}$ & $13.32_{-0.77}^{+0.61}$  \\
		54079-54116 & $57.04_{-0.71}^{+0.85}$ & $9.08_{-0.61}^{+0.77}$ & $21.58_{-2.67}^{+3.87}$ & $7.49_{-0.13}^{+0.19}$ & $11.63_{-0.39}^{+0.29}$  \\
		54116-54215 & $57.89_{-1.09}^{+0.93}$ & $9.40_{-0.58}^{+0.89}$ & $22.63_{-3.53}^{+4.05}$ & $7.24_{-0.19}^{+0.16}$ & $12.85_{-0.53}^{+0.81}$  \\
		54215-54338 & $57.42_{-0.84}^{+1.24}$ & $8.69_{-0.58}^{+0.87}$ & $21.03_{-2.57}^{+3.60}$ & $6.78_{-0.10}^{+0.13}$ & $15.17_{-0.79}^{+1.15}$  \\
		54349-54366 & $56.15_{-0.94}^{+0.58}$ & $7.01_{-1.09}^{+0.36}$ & $14.47_{-2.85}^{+1.13}$ & $6.94_{-0.14}^{+0.10}$ & $13.24_{-0.68}^{+0.73}$  \\
		54366-54420 & $57.69_{-1.71}^{+1.14}$ & $8.94_{-1.01}^{+0.74}$ & $21.54_{-5.73}^{+2.93}$ & $7.23_{-0.21}^{+0.22}$ & $12.41_{-0.69}^{+0.78}$  \\
		54420-54451 & $55.06_{-0.62}^{+0.50}$ & $6.81_{-0.45}^{+0.74}$ & $15.02_{-1.81}^{+1.85}$ & $6.91_{-0.10}^{+0.13}$ & $13.75_{-0.94}^{+0.76}$  \\
		54451-54493 & $55.90_{-1.12}^{+1.14}$ & $8.02_{-0.74}^{+0.46}$ & $15.83_{-2.89}^{+2.36}$ & $6.59_{-0.13}^{+0.10}$ & $15.90_{-1.03}^{+1.13}$  \\
		54494-54519 & $57.40_{-0.64}^{+0.71}$ & $8.25_{-0.46}^{+0.51}$ & $18.19_{-1.64}^{+1.92}$ & $6.78_{-0.05}^{+0.06}$ & $14.80_{-0.22}^{+0.20}$  \\
		54519-54637 & $56.00_{-1.33}^{+0.82}$ & $8.09_{-0.92}^{+0.63}$ & $17.75_{-2.95}^{+3.08}$ & $6.93_{-0.14}^{+0.15}$ & $13.94_{-0.91}^{+0.85}$  \\
		54637-54691 & $55.77_{-0.83}^{+1.07}$ & $8.36_{-0.66}^{+0.75}$ & $17.67_{-2.62}^{+2.90}$ & $6.91_{-0.12}^{+0.13}$ & $14.07_{-0.94}^{+0.65}$  \\
		54691-54769 & $56.71_{-1.21}^{+0.43}$ & $10.30_{-1.06}^{+0.69}$ & $30.25_{-6.79}^{+0.86}$ & $8.05_{-0.40}^{+0.32}$ & $10.27_{-0.48}^{+0.67}$  \\
		54769-54809 & $58.31_{-0.81}^{+0.86}$ & $9.76_{-0.74}^{+0.68}$ & $25.70_{-4.49}^{+4.20}$ & $7.29_{-0.25}^{+0.25}$ & $12.19_{-0.88}^{+0.93}$  \\
		54809-54858 & $57.20_{-0.33}^{+1.35}$ & $9.27_{-0.29}^{+0.74}$ & $19.55_{-1.60}^{+3.39}$ & $6.64_{-0.10}^{+0.13}$ & $16.15_{-0.90}^{+0.99}$  \\
		54858-54871 & $56.97_{-0.82}^{+1.09}$ & $8.76_{-0.48}^{+0.89}$ & $19.64_{-4.34}^{+3.34}$ & $6.67_{-0.12}^{+0.11}$ & $16.83_{-1.14}^{+1.30}$  \\
		54871-54932 & $56.91_{-0.92}^{+1.07}$ & $8.35_{-0.54}^{+0.69}$ & $19.34_{-2.02}^{+3.68}$ & $6.89_{-0.11}^{+0.15}$ & $14.16_{-0.94}^{+0.66}$  \\
		54932-54986 & $59.56_{-1.25}^{+0.46}$ & $9.32_{-0.79}^{+0.54}$ & $21.93_{-3.10}^{+2.89}$ & $6.69_{-0.09}^{+0.10}$ & $15.23_{-0.88}^{+0.58}$  \\
		54986-55068 & $57.71_{-0.99}^{+0.75}$ & $9.99_{-0.77}^{+1.31}$ & $26.63_{-4.05}^{+2.71}$ & $7.40_{-0.16}^{+0.38}$ & $11.84_{-0.58}^{+0.37}$  \\
		55068-55145 & $57.43_{-0.65}^{+0.58}$ & $10.00_{-0.48}^{+0.88}$ & $25.29_{-2.42}^{+1.58}$ & $7.01_{-0.08}^{+0.17}$ & $13.92_{-0.34}^{+0.25}$  \\
		55145-55202 & $55.64_{-0.81}^{+0.92}$ & $8.81_{-0.64}^{+0.75}$ & $17.58_{-2.31}^{+3.06}$ & $7.01_{-0.09}^{+0.12}$ & $13.22_{-0.33}^{+0.28}$  \\
		55240-55311 & $56.35_{-0.72}^{+0.82}$ & $9.14_{-0.71}^{+0.69}$ & $22.45_{-3.56}^{+2.96}$ & $7.11_{-0.20}^{+0.16}$ & $12.94_{-0.82}^{+0.79}$  \\
		55312-55388 & $55.85_{-0.45}^{+0.62}$ & $9.46_{-0.51}^{+0.35}$ & $25.22_{-2.93}^{+2.30}$ & $7.09_{-0.19}^{+0.09}$ & $13.81_{-0.60}^{+1.20}$  \\
		55388-55446 & $57.39_{-0.97}^{+1.23}$ & $8.95_{-0.51}^{+0.91}$ & $22.29_{-2.80}^{+5.00}$ & $7.28_{-0.15}^{+0.24}$ & $12.28_{-0.84}^{+0.59}$  \\
		55447-55522 & $55.45_{-0.72}^{+0.81}$ & $8.18_{-0.54}^{+0.61}$ & $18.28_{-2.02}^{+2.46}$ & $6.89_{-0.07}^{+0.09}$ & $13.95_{-0.28}^{+0.25}$  \\
		55522-55596 & $55.66_{-0.73}^{+0.26}$ & $10.00_{-0.49}^{+0.28}$ & $27.39_{-3.10}^{+1.93}$ & $7.58_{-0.24}^{+0.13}$ & $11.77_{-0.34}^{+0.73}$  \\
		55597-55649 & $53.95_{-0.47}^{+0.69}$ & $8.56_{-0.68}^{+0.56}$ & $18.96_{-3.67}^{+2.64}$ & $7.06_{-0.21}^{+0.11}$ & $12.65_{-0.44}^{+0.74}$  \\
		55650-55739 & $56.00_{-0.82}^{+0.28}$ & $9.99_{-0.67}^{+1.02}$ & $27.89_{-3.91}^{+0.96}$ & $7.31_{-0.17}^{+0.34}$ & $12.17_{-0.46}^{+0.39}$  \\
		55740-55809 & $55.13_{-0.70}^{+0.82}$ & $9.02_{-0.60}^{+0.72}$ & $21.51_{-2.70}^{+3.71}$ & $7.15_{-0.12}^{+0.17}$ & $12.86_{-0.41}^{+0.33}$  \\
		55809-55874 & $54.16_{-0.91}^{+1.08}$ & $8.78_{-0.81}^{+0.94}$ & $18.41_{-3.53}^{+4.87}$ & $6.98_{-0.21}^{+0.27}$ & $13.49_{-1.03}^{+1.18}$  \\
		55874-55933 & $55.14_{-0.70}^{+1.08}$ & $9.80_{-0.62}^{+0.84}$ & $24.14_{-2.78}^{+5.04}$ & $7.49_{-0.14}^{+0.30}$ & $11.61_{-0.72}^{+0.36}$  \\
		55933-55976 & $53.84_{-0.87}^{+1.02}$ & $8.84_{-0.74}^{+0.87}$ & $19.40_{-3.56}^{+4.84}$ & $6.71_{-0.19}^{+0.24}$ & $15.46_{-1.38}^{+1.68}$  \\
		55976-56044 & $53.68_{-0.58}^{+0.75}$ & $8.05_{-0.39}^{+0.70}$ & $15.59_{-2.20}^{+2.27}$ & $6.65_{-0.13}^{+0.13}$ & $15.91_{-0.79}^{+1.30}$  \\
		56044-56126 & $55.83_{-0.71}^{+0.47}$ & $9.96_{-0.61}^{+0.77}$ & $26.86_{-3.41}^{+1.41}$ & $7.10_{-0.14}^{+0.19}$ & $13.60_{-0.35}^{+0.39}$  \\
		56126-56203 & $55.68_{-0.88}^{+1.04}$ & $9.50_{-0.59}^{+0.79}$ & $23.05_{-3.75}^{+4.31}$ & $6.89_{-0.15}^{+0.19}$ & $14.06_{-0.98}^{+0.81}$  \\
		56204-56283 & $55.87_{-0.88}^{+0.42}$ & $8.18_{-0.71}^{+0.53}$ & $16.98_{-2.25}^{+1.76}$ & $6.44_{-0.14}^{+0.08}$ & $17.21_{-0.94}^{+1.46}$  \\
		56283-56375 & $53.12_{-1.07}^{+0.54}$ & $8.12_{-0.76}^{+0.46}$ & $16.30_{-3.50}^{+2.76}$ & $6.79_{-0.21}^{+0.15}$ & $14.78_{-0.91}^{+1.54}$  \\
		56375-56466 & $54.55_{-0.89}^{+1.04}$ & $8.86_{-0.76}^{+0.86}$ & $19.67_{-3.55}^{+4.72}$ & $7.39_{-0.23}^{+0.28}$ & $11.77_{-0.73}^{+0.81}$  \\
		56466-56525 & $55.19_{-0.93}^{+0.53}$ & $10.01_{-0.76}^{+0.94}$ & $25.07_{-4.44}^{+1.68}$ & $6.93_{-0.21}^{+0.04}$ & $14.44_{-0.66}^{+1.27}$  \\
		56525-56591 & $53.49_{-0.92}^{+1.07}$ & $8.64_{-0.80}^{+0.90}$ & $17.82_{-3.41}^{+4.58}$ & $6.95_{-0.20}^{+0.25}$ & $13.57_{-0.99}^{+1.15}$  \\
		56592-56647 & $54.75_{-1.11}^{+0.86}$ & $9.60_{-0.90}^{+1.07}$ & $22.80_{-4.79}^{+3.87}$ & $7.27_{-0.26}^{+0.20}$ & $12.31_{-0.73}^{+0.99}$  \\
		56647-56742 & $53.45_{-0.58}^{+0.99}$ & $8.52_{-0.44}^{+0.89}$ & $16.98_{-2.87}^{+2.38}$ & $6.72_{-0.17}^{+0.16}$ & $15.06_{-1.06}^{+1.11}$  \\
		56742-56839 & $55.88_{-1.28}^{+0.99}$ & $10.00_{-0.74}^{+1.03}$ & $23.96_{-5.29}^{+4.93}$ & $6.96_{-0.23}^{+0.22}$ & $14.18_{-0.82}^{+1.12}$  \\
		56839-56879 & $53.35_{-0.50}^{+0.56}$ & $8.23_{-0.43}^{+0.48}$ & $18.22_{-1.69}^{+2.05}$ & $6.81_{-0.08}^{+0.09}$ & $14.23_{-0.29}^{+0.27}$  \\
		56880-56951 & $55.98_{-0.87}^{+0.71}$ & $9.99_{-0.71}^{+1.63}$ & $23.34_{-3.34}^{+1.03}$ & $7.06_{-0.13}^{+0.35}$ & $13.32_{-0.46}^{+0.36}$  \\
		56951-57003 & $55.95_{-1.15}^{+0.43}$ & $10.00_{-0.74}^{+0.45}$ & $21.34_{-3.52}^{+3.24}$ & $6.78_{-0.15}^{+0.19}$ & $15.74_{-1.15}^{+1.27}$  \\
		57003-57040 & $54.09_{-0.68}^{+0.92}$ & $9.54_{-0.67}^{+0.77}$ & $20.51_{-3.68}^{+2.72}$ & $7.27_{-0.21}^{+0.17}$ & $12.31_{-0.65}^{+0.80}$  \\
		57040-57085 & $54.51_{-0.98}^{+0.85}$ & $9.67_{-0.80}^{+0.91}$ & $26.24_{-4.91}^{+2.62}$ & $7.24_{-0.26}^{+0.14}$ & $12.51_{-0.75}^{+1.01}$  \\
		57085-57197 & $55.66_{-1.23}^{+0.44}$ & $10.76_{-1.07}^{+0.56}$ & $31.56_{-7.02}^{+1.02}$ & $7.59_{-0.32}^{+0.21}$ & $12.09_{-0.46}^{+0.70}$  \\
		57197-57235 & $55.36_{-1.18}^{+0.59}$ & $8.14_{-0.73}^{+0.46}$ & $19.24_{-3.39}^{+2.83}$ & $7.30_{-0.20}^{+0.23}$ & $12.29_{-0.89}^{+0.70}$  \\
		57235-57298 & $55.51_{-1.22}^{+1.10}$ & $8.47_{-0.90}^{+0.89}$ & $19.48_{-3.44}^{+3.84}$ & $7.14_{-0.21}^{+0.19}$ & $13.11_{-0.80}^{+1.08}$  \\
		57298-57386 & $55.79_{-0.57}^{+0.78}$ & $8.64_{-0.28}^{+0.88}$ & $23.03_{-2.30}^{+4.83}$ & $7.52_{-0.10}^{+0.20}$ & $11.50_{-0.65}^{+0.46}$  \\
		57386-57458 & $55.79_{-0.68}^{+0.83}$ & $9.68_{-0.81}^{+0.67}$ & $24.54_{-3.42}^{+3.68}$ & $7.18_{-0.25}^{+0.14}$ & $13.30_{-0.62}^{+1.07}$  \\
		57459-57506 & $54.31_{-0.67}^{+0.74}$ & $7.35_{-0.47}^{+0.52}$ & $16.14_{-1.64}^{+1.90}$ & $6.61_{-0.05}^{+0.06}$ & $16.63_{-0.28}^{+0.27}$  \\
		57506-57560 & $55.12_{-0.56}^{+1.23}$ & $8.55_{-0.44}^{+1.07}$ & $23.06_{-2.45}^{+4.65}$ & $7.05_{-0.20}^{+0.17}$ & $13.59_{-0.83}^{+0.85}$  \\
		57560-57610 & $55.88_{-0.47}^{+1.40}$ & $9.02_{-0.49}^{+0.97}$ & $21.97_{-3.05}^{+5.02}$ & $7.11_{-0.22}^{+0.21}$ & $12.59_{-0.78}^{+0.92}$  \\
		57610-57687 & $53.56_{-0.48}^{+0.91}$ & $7.60_{-0.44}^{+0.76}$ & $15.55_{-2.33}^{+3.25}$ & $6.72_{-0.14}^{+0.17}$ & $15.43_{-1.20}^{+1.12}$  \\
		57687-57765 & $54.94_{-0.73}^{+1.10}$ & $8.50_{-0.43}^{+0.76}$ & $19.22_{-3.69}^{+4.44}$ & $6.97_{-0.20}^{+0.18}$ & $13.76_{-0.86}^{+0.94}$  \\
		57765-57800 & $53.77_{-0.46}^{+0.78}$ & $6.99_{-0.31}^{+0.64}$ & $14.30_{-1.36}^{+2.14}$ & $6.74_{-0.07}^{+0.15}$ & $14.85_{-1.11}^{+0.61}$  \\
		57821-57898 & $56.19_{-0.83}^{+0.50}$ & $9.00_{-0.56}^{+0.45}$ & $22.75_{-4.43}^{+2.85}$ & $7.09_{-0.19}^{+0.19}$ & $13.42_{-0.82}^{+1.03}$  \\
		57898-57969 & $57.38_{-1.02}^{+1.01}$ & $9.20_{-0.84}^{+0.58}$ & $23.64_{-4.43}^{+5.23}$ & $6.97_{-0.16}^{+0.20}$ & $13.79_{-0.96}^{+0.84}$  \\
		57969-58017 & $56.84_{-1.39}^{+1.26}$ & $8.47_{-1.08}^{+0.89}$ & $17.44_{-3.47}^{+2.77}$ & $6.45_{-0.11}^{+0.09}$ & $19.17_{-1.32}^{+1.90}$  \\
		58018-58108 & $55.46_{-0.84}^{+1.39}$ & $7.80_{-0.47}^{+1.12}$ & $17.33_{-2.48}^{+3.87}$ & $6.75_{-0.12}^{+0.17}$ & $14.74_{-1.13}^{+0.96}$  \\
		58108-58181 & $54.52_{-0.96}^{+0.59}$ & $7.33_{-1.03}^{+0.54}$ & $16.23_{-3.31}^{+2.59}$ & $6.84_{-0.14}^{+0.12}$ & $14.25_{-0.89}^{+0.73}$  \\
		58181-58260 & $55.05_{-1.33}^{+0.69}$ & $8.36_{-1.11}^{+0.82}$ & $17.86_{-4.21}^{+3.24}$ & $6.83_{-0.18}^{+0.17}$ & $14.35_{-1.01}^{+0.93}$  \\
		\hline
%\caption{Best-fitting parameters of Vela X-1 observed with {\it Swift}/BAT. $E_{\rm cyc\_H}$\,(keV), $\sigma_{\rm H}\,(keV)$ and Depth$_{\rm H}$ are the energy, width and depth of the harmonic CRSF, respectively.  $kT_{\rm Comp}$ and $\tau_{\rm Comp}$ are the temperature and the optical depth of the Comptonized continuum spectra, respectively.}
\end{longtable}
\clearpage
\begin{longtable}{ c|ccccc }
\caption{Best-fitting parameters of Cen X-3 observed with {\it Swift}/BAT. $E_{\rm cyc}$, $\sigma$ and Depth are the energy, width and depth of the CRSF.
The continuum model is a powerlaw  modified by a Fermi-Dirac cutoff, where $\Gamma$ and  $E_{\rm fold}$ are the powerlaw index and the e-folding parameter.
}
\label{tab_cen}\\
		\hline
		Time (MJD) & $E_{\rm cyc}$\,(keV) & $\sigma\,(keV)$ & Depth & $\Gamma$ &  $E_{\rm fold}$\\
		\hline
		53422-53595 & $32.18_{-0.46}^{+0.30}$ & $7.17_{-0.84}^{+0.70}$ & $12.44_{-2.17}^{+2.37}$ &  	$1.35_{-0.04}^{+0.07}$ & $8.55_{-0.26}^{+0.39}$  \\
		53596-53707 & $31.70_{-0.94}^{+2.07}$ & $7.73_{-1.39}^{+2.42}$ & $11.88_{-4.95}^{+9.29}$ &  	$1.00_{-0.13}^{+0.19}$ & $6.85_{-0.33}^{+0.62}$  \\
		53717-53889 & $31.61_{-1.41}^{+1.05}$ & $5.35_{-1.17}^{+1.17}$ & $5.87_{-2.11}^{+2.06}$ &  	$1.84_{-0.13}^{+0.16}$ & $8.72_{-0.13}^{+0.53}$  \\
		53897-53955 & $31.67_{-0.61}^{+2.01}$ & $4.36_{-0.63}^{+1.68}$ & $4.80_{-0.95}^{+3.44}$ &  	$1.76_{-0.12}^{+0.15}$ & $8.10_{-0.31}^{+0.78}$  \\
		53956-54015 & $32.66_{-1.73}^{+1.61}$ & $8.69_{-1.52}^{+0.97}$ & $15.46_{-8.04}^{+3.64}$ &  	$1.11_{-0.14}^{+0.18}$ & $7.22_{-0.23}^{+0.84}$  \\
		54022-54071 & $30.24_{-0.88}^{+0.55}$ & $2.51_{-1.19}^{+0.85}$ & $1.85_{-0.84}^{+0.40}$ &  	$2.60_{-0.02}^{+0.05}$ & $12.96_{-0.60}^{+1.50}$  \\
		54076-54135 & $30.69_{-1.70}^{+0.87}$ & $9.96_{-0.93}^{+1.98}$ & $18.50_{-4.72}^{+5.13}$ &  	$0.95_{-0.11}^{+0.16}$ & $7.14_{-0.35}^{+0.56}$  \\
		54136-54195 & $30.26_{-0.38}^{+1.09}$ & $6.32_{-0.62}^{+0.65}$ & $8.19_{-1.56}^{+1.23}$ &  	$1.49_{-0.09}^{+0.11}$ & $8.25_{-0.17}^{+0.33}$  \\
		54230-54366 & $31.94_{-1.44}^{+0.60}$ & $8.04_{-1.54}^{+3.77}$ & $14.14_{-7.06}^{+9.64}$ &  	$0.93_{-0.03}^{+0.12}$ & $7.36_{-0.17}^{+0.61}$  \\
		54403-54495 & $30.18_{-1.05}^{+2.35}$ & $6.22_{-0.12}^{+2.25}$ & $9.02_{-0.71}^{+5.67}$ &  	$0.98_{-0.04}^{+0.05}$ & $7.54_{-0.79}^{+1.01}$  \\
		54496-54545 & $34.69_{-1.39}^{+1.64}$ & $9.99_{-1.31}^{+1.49}$ & $26.18_{-11.52}^{+7.61}$ &  	$0.90_{-0.10}^{+0.17}$ & $7.81_{-0.23}^{+0.93}$  \\
		54558-54675 & $32.13_{-0.76}^{+0.45}$ & $6.10_{-0.73}^{+0.24}$ & $8.08_{-2.25}^{+0.62}$ &  	$1.72_{-0.01}^{+0.03}$ & $8.79_{-0.30}^{+0.60}$  \\
		54679-54777 & $33.40_{-1.22}^{+3.27}$ & $9.98_{-1.83}^{+4.13}$ & $20.20_{-10.62}^{+17.24}$ &  	$0.86_{-0.08}^{+0.16}$ & $7.17_{-0.20}^{+0.92}$  \\
		54800-54854 & $30.55_{-2.50}^{+2.99}$ & $9.95_{-0.37}^{+0.60}$ & $15.90_{-7.90}^{+5.83}$ &  	$0.92_{-0.17}^{+0.24}$ & $6.94_{-0.35}^{+0.87}$  \\
		54087-54975 & $31.84_{-0.66}^{+0.90}$ & $5.41_{-1.02}^{+0.82}$ & $5.66_{-1.35}^{+0.88}$ &  	$1.86_{-0.13}^{+0.15}$ & $9.52_{-0.30}^{+0.49}$  \\
		54986-55151 & $30.41_{-0.82}^{+2.36}$ & $9.98_{-3.01}^{+5.32}$ & $17.45_{-11.09}^{+22.57}$ &  	$0.93_{-0.13}^{+0.23}$ & $7.34_{-0.37}^{+0.54}$  \\
		55161-55215 & $29.53_{-0.40}^{+0.71}$ & $3.37_{-0.53}^{+0.33}$ & $3.02_{-0.45}^{+0.55}$ &  	$2.22_{-0.11}^{+0.13}$ & $11.00_{-0.40}^{+0.63}$  \\
		55216-55272 & $32.45_{-0.81}^{+1.69}$ & $9.62_{-1.84}^{+3.17}$ & $18.29_{-8.05}^{+12.77}$ &  	$0.81_{-0.03}^{+0.12}$ & $6.98_{-0.32}^{+0.52}$  \\
		55278-55452 & $31.37_{-0.72}^{+0.42}$ & $6.19_{-0.96}^{+0.67}$ & $8.10_{-2.17}^{+1.14}$ &  	$1.57_{-0.04}^{+0.08}$ & $8.55_{-0.41}^{+0.53}$  \\
		55465-55515 & $32.85_{-0.59}^{+0.96}$ & $7.69_{-1.19}^{+1.48}$ & $15.18_{-3.71}^{+3.66}$ &  	$1.28_{-0.03}^{+0.08}$ & $8.20_{-0.30}^{+0.40}$  \\
		55517-55631 & $34.02_{-1.14}^{+1.95}$ & $10.02_{-1.81}^{+1.57}$ & $23.03_{-10.27}^{+9.37}$ &  	$0.94_{-0.10}^{+0.19}$ & $8.04_{-0.29}^{+1.03}$  \\
		55638-55804 & $31.68_{-0.74}^{+1.71}$ & $9.52_{-1.92}^{+3.10}$ & $16.99_{-8.80}^{+10.86}$ &  	$1.20_{-0.09}^{+0.15}$ & $7.92_{-0.30}^{+0.69}$  \\
		55833-55875 & $31.75_{-0.65}^{+0.75}$ & $6.33_{-1.52}^{+2.91}$ & $7.99_{-3.36}^{+12.19}$ &  	$1.30_{-0.10}^{+0.12}$ & $7.74_{-0.13}^{+0.38}$  \\
		55876-55935 & $32.68_{-0.51}^{+0.56}$ & $9.99_{-1.15}^{+0.85}$ & $20.19_{-6.31}^{+2.80}$ &  	$1.16_{-0.06}^{+0.12}$ & $7.67_{-0.36}^{+0.44}$  \\
		%55936-55995 & $28.58_{--0.03}^{+0.71}$ & $10.00_{-1.16}^{+0.30}$ & $33.54_{-9.46}^{+1.30}$ &  	$-1.87_{-0.14}^{+0.19}$ & $4.16_{-0.13}^{+0.17}$  \\
		55996-56055 & $30.92_{-0.69}^{+0.34}$ & $5.48_{-1.17}^{+0.51}$ & $6.01_{-2.12}^{+0.80}$ &  	$1.44_{-0.13}^{+0.16}$ & $7.77_{-0.16}^{+0.48}$  \\
		56056-56163 & $31.71_{-1.23}^{+0.36}$ & $6.13_{-1.09}^{+0.32}$ & $7.40_{-2.85}^{+0.44}$ &  	$1.58_{-0.12}^{+0.14}$ & $8.57_{-0.25}^{+0.67}$  \\
		56207-56278 & $31.49_{-1.03}^{+2.20}$ & $10.00_{-1.39}^{+3.98}$ & $17.24_{-4.24}^{+14.38}$ &  	$0.83_{-0.02}^{+0.09}$ & $6.96_{-0.20}^{+0.46}$  \\
		56307-56355 & $31.84_{-0.49}^{+0.75}$ & $7.30_{-0.60}^{+0.37}$ & $10.87_{-2.05}^{+1.60}$ &  	$1.35_{-0.02}^{+0.02}$ & $7.96_{-0.22}^{+0.38}$  \\
		56356-56413 & $32.07_{-0.54}^{+0.49}$ & $6.62_{-0.81}^{+0.42}$ & $8.33_{-1.40}^{+0.46}$ &  	$1.66_{-0.13}^{+0.14}$ & $8.49_{-0.24}^{+0.38}$  \\
		56416-56595 & $31.85_{-1.17}^{+0.52}$ & $7.19_{-1.04}^{+0.45}$ & $11.47_{-3.15}^{+0.65}$ &  	$0.69_{-0.09}^{+0.12}$ & $6.83_{-0.16}^{+0.37}$  \\
		56596-56631 & $32.50_{-0.83}^{+2.41}$ & $9.37_{-1.65}^{+1.95}$ & $17.95_{-8.29}^{+10.68}$ &  	$1.22_{-0.13}^{+0.17}$ & $7.65_{-0.28}^{+0.84}$  \\
		56702-56835 & $31.78_{-1.02}^{+0.82}$ & $9.58_{-1.71}^{+1.48}$ & $16.70_{-6.88}^{+5.51}$ &  	$0.98_{-0.07}^{+0.08}$ & $7.03_{-0.15}^{+0.29}$  \\
		56838-57008 & $32.42_{-1.17}^{+2.02}$ & $10.00_{-2.07}^{+3.05}$ & $18.32_{-9.06}^{+11.84}$ &  	$0.74_{-0.07}^{+0.12}$ & $6.94_{-0.28}^{+0.60}$  \\
		57020-57067 & $31.62_{-0.45}^{+1.02}$ & $8.78_{-1.26}^{+0.32}$ & $14.55_{-3.79}^{+1.24}$ &  	$0.97_{-0.09}^{+0.11}$ & $7.08_{-0.18}^{+0.32}$  \\
		57102-57246 & $31.37_{-0.81}^{+0.60}$ & $6.74_{-1.47}^{+0.94}$ & $9.04_{-3.56}^{+2.39}$ &  	$1.50_{-0.11}^{+0.14}$ & $8.73_{-0.24}^{+0.38}$  \\
		57276-57359 & $32.65_{-0.67}^{+0.87}$ & $9.14_{-1.47}^{+1.62}$ & $17.65_{-7.07}^{+5.56}$ &  	$1.34_{-0.15}^{+0.18}$ & $8.51_{-0.22}^{+0.59}$  \\
		57379-57405 & $32.81_{-0.90}^{+1.47}$ & $10.00_{-1.20}^{+0.23}$ & $19.82_{-3.95}^{+1.78}$ &  	$1.16_{-0.04}^{+0.08}$ & $8.14_{-0.41}^{+0.60}$  \\
		57436-57615 & $29.75_{-0.25}^{+0.53}$ & $6.35_{-0.60}^{+1.07}$ & $6.86_{-0.99}^{+1.98}$ &  	$1.22_{-0.14}^{+0.16}$ & $7.38_{-0.25}^{+0.41}$  \\
		57616-57735 & $30.79_{-0.61}^{+1.31}$ & $9.32_{-1.67}^{+2.76}$ & $15.12_{-4.98}^{+8.55}$ &  	$1.31_{-0.13}^{+0.17}$ & $8.09_{-0.31}^{+0.65}$  \\
		57736-57795 & $28.89_{-1.45}^{+1.42}$ & $10.00_{-1.00}^{+1.37}$ & $14.38_{-4.25}^{+4.97}$ &  	$0.71_{-0.18}^{+0.27}$ & $6.25_{-0.44}^{+0.70}$  \\
		57614-57851 & $31.65_{-1.05}^{+0.79}$ & $10.02_{-2.82}^{+3.67}$ & $14.44_{-7.35}^{+10.37}$ &  	$0.69_{-0.11}^{+0.20}$ & $6.48_{-0.25}^{+0.60}$  \\
		57856-57915 & $31.44_{-0.33}^{+0.27}$ & $7.09_{-1.04}^{+0.75}$ & $11.03_{-3.07}^{+1.87}$ &  	$1.34_{-0.04}^{+0.09}$ & $8.27_{-0.30}^{+0.39}$  \\
		57916-57956 & $33.31_{-0.94}^{+0.96}$ & $9.26_{-1.09}^{+1.78}$ & $16.41_{-5.46}^{+6.83}$ &  	$1.10_{-0.10}^{+0.15}$ & $7.54_{-0.23}^{+0.41}$  \\
		57984-58009 & $32.44_{-1.33}^{+1.39}$ & $8.51_{-0.77}^{+0.61}$ & $14.50_{-4.60}^{+4.45}$ &  	$1.03_{-0.10}^{+0.13}$ & $7.39_{-0.25}^{+0.63}$  \\
		58037-58155 & $33.12_{-1.61}^{+1.88}$ & $9.96_{-1.41}^{+2.33}$ & $19.24_{-7.89}^{+8.02}$ &  	$1.12_{-0.03}^{+0.07}$ & $7.91_{-0.44}^{+0.85}$  \\
		58159-58267 & $34.31_{-0.58}^{+0.90}$ & $8.28_{-1.37}^{+0.50}$ & $16.14_{-4.72}^{+1.78}$ &  	$1.27_{-0.13}^{+0.16}$ & $8.24_{-0.36}^{+0.67}$  \\
\hline
\end{longtable}
%\end{comment}

%%%%%%%%%%%%%%%%%%%%%%%%%%%%%%%%%%%%%%%%%%%%%%%%%%

%%%%%%%%%%%%%%%%% APPENDICES %%%%%%%%%%%%%%%%%%%%%

\clearpage

% Don't change these lines
\bsp	% typesetting comment
\label{lastpage}
\end{document}